# Optical properties of a diamond NV color center from capped embedded multiconfigurational correlated wavefunction theory

John Mark P. Martirez
Applied Materials and Sustainability Sciences, Princeton Plasma Physics Laboratory,
Princeton, New Jersey, 08540
martirez@pppl.gov, martirez@princeton.edu

**Abstract**

Diamond defects are among the most promising qubits. Modelling their properties through accurate quantum mechanical simulations can further their development into robust units of information. We use the recently developed capped density functional embedding theory (capped-DFET) with the multiconfigurational *n*-electron valence second-order perturbation theory to characterize the electronic excitation energies for different spin manifolds of the well-characterized negatively charged substitutional N defect adjacent to a vacancy ($V_C$) in diamond ($N_CV_C^-$). We successfully reproduce vertical excitation energies for both triplet and singlet states of $N_CV_C^-$ with errors < 0.1 eV. Unlike other embedding methods, capped-DFET exhibits robust predictions that are approximately independent of the embedded cluster size: it only requires a cluster to contain the defect atoms and their nearest neighbors (as small as a 40-atom capped cluster). Furthermore, our method is free from slowly converging Coulomb interactions between charged defects, and thus also only weakly dependent on supercell size.

**Introduction**

Quantum coherence, e.g., of electron and nuclear spins, is key to establishing quantum bits (qubits) suitable for quantum computing.[1-3] Quantum coherence in qubits is the core property that classical bits do not possess, in that the written value of a classical bit can only be either "0" or "1" at a given moment, whereas a quantum bit can be a superposition of both "0" and "1" states, only resolving its value upon measurement.[3] Qubits can be made from isolated atoms or defects in a material, whose spin and charge are well isolated from its host, so as not to interfere with qubit preparation, manipulation, and read-out.[1-3] Quantum coherence is not only potentially revolutionary for computing, but also for communication, data security, and sensing.[4]

Defect properties, such as spin, charge, and optical response, are dictated by: the atoms that compose them; their local environment, e.g., coordination number, identity of neighbors, strain, etc.; and, especially in the case of charged defects, the presence of other (charged) defects. A quantum defect host meant for quantum computing applications, ideally, has a wide band gap (so that the host does not absorb the light used to probe the defects), can be made with high purity, is diamagnetic, induces small spin-orbit coupling, and has naturally occurring isotopes with no nuclear spin.[3,5] Diamond is among the most promising (and therefore, among most studied) materials for solid-state quantum applications because it can host an array of magnetically (via electron spin) and optically active point defects, namely, vacancies, substitutions (e.g., H, N, Si, B, O, Ge, P, Sn, Pb, etc.), interstitials, and combinations thereof that are either neutral or positively or negatively charged.[6,7] Take, for example, a well-known solid-state qubit: the negatively charged substitutional N defect ($N_C$) with an adjacent vacancy ($V_C$) in diamond ($N_CV_C^-$). The removal of two adjacent C atoms from the diamond lattice creates six C atoms with only three neighbors, rendering them coordinatively saturated (leaves a "dangling bond" per C) with nominally one unpaired electron ("open shell") on each. However, when an N atom occupies one of the vacancy sites, three of the six C atoms form bonds with the N and the N, having one more electron than C, will be coordinatively saturated ("closed shell"), forming three covalent bonds and a lone pair of electrons. This leaves three open shell C atoms, each with one ($sp^3$) orbital containing only one electron (**Fig. 1A**). This defect is called an $NV^0$ center. If this defect is negatively charged ($N_CV_C^-$), the additional electron can pair with one of the three unpaired $sp^3$ electrons (**Fig. 1A**). Indeed, in the ground state, $NV^-$ center has two unpaired electrons – a spin triplet.[8] **Fig. 1B** shows the electronic configuration of the spin triplet ground- and first excited-state states labeled $^3A_2$ and $^3E$ (the superscript 3 refers to their spin multiplicity and the alphanumeric labels originate from the defect symmetry – *vide infra*), respectively, showing how the three C $sp^3$-derived states at the vacancy are populated (one lower energy and two degenerate higher energy orbitals).

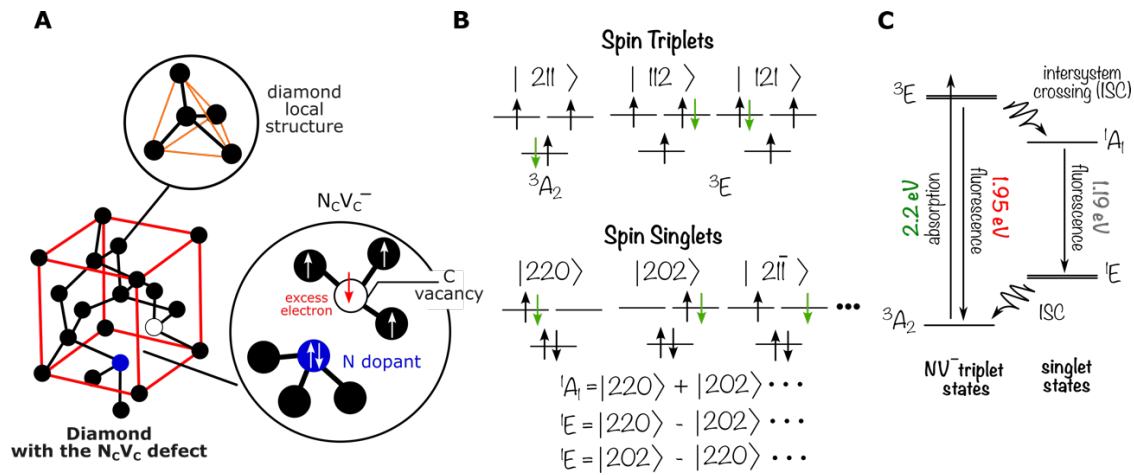

**Fig. 1**. **Atomic and electronic structures of a prototypical defect in diamond**. **A**, right: diamond is composed of C atoms each with four nearest neighbors that are ~1.5 Å away in a tetrahedral local geometry (top inset). The cubic diamond unit cell contains eight unique C atoms with cell side lengths ~3.6 Å (red box). Left: an NV center consists of a N substitutional defect adjacent to a C vacancy. **B**, $N_CV_C^-$ center has a triplet spin ground state (i.e., two unpaired electrons). Electron occupation of the three orbitals derived from the dangling $sp^3$ orbitals are shown for both triplet ($^3A_2$ and $^3E$) and singlet states ($^1E$ and $^1A_1$). For the spin singlet, only three of the six possible electronic configurations are shown. The bra-ket notation, e.g., |211⟩, refers to the wavefunction associated with a specific occupation of the three

defect orbitals: either doubly (2), spin-up singly (1), or spin-down singly ($\bar{1}$) occupied or unoccupied (0). **C**, diamond absorbs in the deep UV (5.5 eV), whereas the ground-state spin triplet NV⁻ center absorbs and fluoresces within the visible (~2.0 eV) - hence referred to as a "color center". The energy diagram schematically shows the relative energetic positions of the four electronic states in **B** and the various optical processes that link them.

In the $^3A_2 \rightarrow {}^3E$ excitation, a spin-down electron (**Fig. 1B**, green pointing down arrow) in the lowest energy frontier orbital gets excited to one of the two singly occupied highest energy orbitals. While $^3A_2$ is singly degenerate, $^3E$ is doubly degenerate due to the two energetically equivalent ways of adding an excited electron to the two (degenerate) higher energy orbitals. Moreover, higher-energy spin singlets result from low-spin pairing of all the electrons. Among the primary electron configurations are where one of the two highest defect frontier orbitals is empty – there are two such configurations, and where the degenerate states are singly occupied but of opposite spins (**Fig. 1B**). These are only three of the six possible spin singlet configurations (*vide infra*). The spin singlet ground- and first excited-states are labeled $^1E$ (doubly degenerate) and $^1A_1$ (non-degenerate), respectively. The $^1E$ states are "destructive" combinations of the $|220\rangle$ and $|202\rangle$ wavefunctions (and some other terms), whereas the $^1A_1$ state is a "constructive" combination of the aforementioned wavefunctions (and some other contributions). **Fig. 1C** shows an energy diagram for all four states and the electronic processes that connect them: absorption, fluorescence, and intersystem crossing (ISC). The excitation and fluorescence energies for $^3A_2 \rightarrow {}^3E$ (2.2 and 1.95 eV, visible)[9] and $^1E \rightarrow {}^1A_1$ (1.19 eV, infrared)[10] are much smaller than diamond's band gap (5.5 eV, ultraviolet).[11]

By assigning Boolean logic elements "1" and "0" to different spin magnetic quantum states ($m_s$) sub-levels, (a qubit of) information can be "written." The relative energy of the different $m_s$ of the spin triplet can be manipulated, e.g., with a static magnetic field (along the defect axis) where parallel: $m_s = +1$ ($|211\rangle$) and anti-parallel: $m_s = -1$ ($|2\bar{1}\bar{1}\rangle$) spins have splitting energy proportional to the external magnetic field strength, with the latter being higher in energy (Zeeman shift). Under zero field, the splitting between the $m_s = 0$ ($|21\bar{1}\rangle$) and $m_s = \pm 1$ sub-levels (zero-field splitting) are in the microwave region (ground-state zero-field magnetic resonance of 2.88 GHz).[12] Therefore, microwave pulses can be used to manipulate coherence between the spin states that are then optically probed (read out) via the $^3A_2 \rightarrow {}^3E$ excitation and its subsequent spin-dependent fluorescence (emission).[13] Spin-forbidden relaxation via ISC of triplets to singlets, e.g., from $^3E$ to the non-magnetic $^1A_1$ state (**Fig. 1C**) will render the qubit magnetically inactive ("shelving").[14]

Fundamental research opportunities abound in diamond-based quantum materials, notably establishing rules for discovery of viable new defects and alternative host materials is in its infancy.[1,2,5,7,15] The quantum mechanical (QM) aspect of the mode of operation of qubits (and multiconfigurational nature of the defect electronic states) motivates the importance of using accurate multireference QM theories to predict the electronic structure of qubits considering different electronic spins and charge states.

Understanding these fundamental properties will be key to qubit characterization, operation, design, and eventually optimization. QM simulations within accurate electronic structure theory to understand and characterize the magnetic and optical properties of existing and candidate defects would be valuable for addressing these challenges. So far, QM simulations of qubits have been performed largely within (single-determinant, i.e., single-configuration) density functional theory (DFT)[16,17] and studies that used excited-state theories to understand them have only been conducted for a few known defects, notably $N_CV_C^-$.[18-24] In cases where experiments are available, comparisons between DFT simulations and experiments are at best approximate and incomplete, although DFT may offer a qualitative picture on the nature of the magnetic and optical properties of a qubit.[25] When exploring a new phase space for defects and hosts, a more rigorous electronic structure theory would be required for a reliable prediction to be made to ascertain that novel systems are worth experimental investigation.

Here, we computationally characterize the electronic structure of the $N_CV_C^-$ color center via capped density functional embedding theory (capped-DFET)[26] and for the first time test the capped-DFET method to characterize localized defects in condensed matter. Specifically, the goal of this study is to test the ability of this method to reproduce experimental magnetic and electronic properties of the $N_CV_C^-$ defect when combined with a rigorous multiconfigurational correlated wavefunction theory, namely, the complete active space self-consistent field (CASSCF) theory[27,28] to incorporate *static electron correlation*, together with a multireference second order perturbation (MRPT2) method, specifically, the *n*-electron valence second-order perturbation theory (NEVPT2),[29] to include *dynamic electron correlation*. Static correlation is particularly important to accurately describe electronic state degeneracies as well as excited states.

**Theory and Methods**

*Capped-DFET theory*. In standard DFET by Huang, Pavone, and Carter, one divides the system into fragments, typically, into two: the region of interest (typically, a small cluster) and the remaining atoms (cluster's environment).[30] One then obtains the inter-fragment interaction via the optimized effective potential (OEP) method within DFT, where one maximizes the extended Wu-Yang functional[31]

$$W = E_{DFT}^{cl}[\rho^{cl}, V_{emb}] + E_{DFT}^{env}[\rho^{env}, V_{emb}] - \int V_{emb} \rho^{full} dr \qquad \text{Eq. 1}$$

The above equation is a constrained maximization of the sum of the fragment DFT energies ($E_{DFT}^{cl}[\rho^{cl}, V_{emb}] + E_{DFT}^{env}[\rho^{env}, V_{emb}]$), where, $cl, env$, and $full$ refer to the cluster, environment, and the full system. The embedding potential ($V_{emb}$) is the Lagrange multiplier to be variationally optimized and the electron densities ($\rho$) define the gradient:

$$\frac{\delta W}{\delta V_{emb}(r)} = \frac{\delta E_{DFT}^{cl}}{\delta V_{emb}(r)} + \frac{\delta E_{DFT}^{env}}{\delta V_{emb}(r)} - \frac{\delta\{\int V_{emb}(r) \rho^{full} dr\}}{\delta V_{emb}(r)} \qquad \text{Eq. 2}$$

$$\frac{\delta W}{\delta V_{emb}(r)} = \rho^{cl}(r) + \rho^{env}(r) - \rho^{full}(r) \qquad \textbf{Eq. 3}$$

with $\rho^{cl}(r)$ and $\rho^{env}(r)$ respectively the self-consistent densities of the cluster and environment fragments, whereas $\rho^{full}(r)$ is the self-consistent density of the full system. At maximum, naturally, **Eq. 3** equals 0 for all $r$. In other words, the action of the optimized local potential $V_{emb}(r)$ on the cluster and environment leads to the reproduction of the full electron density from the sum of their densities. DFET in conjunction with correlated wavefunction methods has been used extensively to simulate reactions on metal surfaces and in solution both in ground (e.g., thermal[32-39] and electrochemical[40-43]) and electronically excited (e.g., plasmonic catalysis[44,45]) states. Various QM-based partitioning and embedding methods exist in the literature that either relies on density or orbital partitioning. For a recent review on the subject, see ref. [46].

When a chemical system is fragmented, electrons are also partitioned along with the atoms, in doing so, a user assumes how electrons distribute among the fragments. When partitioning across metallic, ionic, and non-covalent (hydrogen or van der Waals) bonds, both the charge and atom partitioning are rather straightforward. The *sp*-valence electrons are free-electron-like for metals, and ions and molecules generally are closed shell therefore fragments may simply be assigned electrons according to the total number of electrons of their constituent elements. For covalent bonds, following the same rule (method 1) will likely lead to open-shell fragments; if one considers the polarity of the bond, one can also heterolytically break the bonds, where one assigns all shared electrons between a bond to one fragment (method 2).[47,48] Alternatively, one can partition using option 1 and "cap" the atoms at the fragmentation points using elements of complementary valence (method 3), e.g., a broken single bond will be capped with a monovalent element. Martirez and Carter subsequently introduced a modification of DFET[26] following method 3 to make the method suitable for when covalent or ionic bonds are broken due to partitioning by introducing "capping" elements - akin to the link-atom method implemented in QM/Molecular Mechanics (QM/MM) methods[49] typically used to model large organic and bio-organic molecules. It is therefore similar in spirit to method 2 charge partitioning, without *a priori* input from the user on how charge is to be distributed across bonds, granted a valence complementary atom is used to cap the dangling bonds. In so-called capped-DFET, Martirez and Carter proposed that in addition to the introduction of capping elements to all fragments, an auxiliary fragment composed of all capping elements may be used to project out the densities of the capping elements, which are not native to the original system.[26] In doing so, a completely local representation of the $V_{emb}(r)$ is preserved, charge partitioning is simplified, and most importantly, use of projector orbitals or implementation of an orbital localization scheme is avoided. **Eq. 1** is thus rewritten as

$$W = E_{DFT}^{cl+cap1}[\rho^{cl}, V_{emb}] + E_{DFT}^{env+cap2}[\rho^{env}, V_{emb}] - \int V_{emb} (\rho^{full} + \rho^{cap1+cap2}) \, dr \quad \text{Eq. 4}$$

**Eq. 3** (the gradient) is thus rewritten as:

$$\frac{\delta W}{\delta V_{emb}(r)} = \rho^{cl+cap1}(r) + \rho^{env+cap2}(r) - \rho^{full}(r) - \rho^{cap1+cap2} \quad \text{Eq. 5}$$

where $\rho^{cl+cap1}(r)$, $\rho^{env+cap2}(r)$, and $\rho^{cap1+cap2}$ are the self-consistent electron densities of the cluster with capping elements 1 ($cap1$), environment with capping elements 2 ($cap2$), and the combined capping elements $cap1 + cap2$, i.e., the auxiliary fragment. $V_{emb}(r)$ therefore furnishes the missing influence of the environment on the cluster and of the cluster on the environment not provided by their respective capping elements.

DFET and capped-DFET can enable regional accurate treatment of the cluster, e.g., an accurate correlated wavefunction (CW) description locally, while also incorporating the influence of its environment quantum mechanically via the optimized $V_{emb}$. We refer to such a method as capped embedded CW or capped emb-CW theory. Further discussion on the nature of the approximation and implementation of capped-DFET can be found in Ref. [26]. Emb-CW and capped emb-CW are used to provide a regional (within the cluster) correction to the total energy of the chemical system, enabling the calculation of reaction energies and barriers or even full energy landscapes for electronic ground and excited states, the latter being beyond the ability of time-independent DFT. When evaluating excited-state energies using emb-CW, one approximates the influence of the environment as fixed and the excited state as local: a strongly coupled electron-hole pair (exciton) is generated and only the orbitals of the cluster atoms are involved. Thus, only localized excited states will be properly described. For a fixed number of electrons, the excitation energy from the ground ($n = 0$) to the $n$th excited state within capped emb-CW is simply

$$\Delta E(0 \to n) = E_{CW}^{cl+cap1}(n) - E_{CW}^{cl+cap1}(n = 0) \quad \text{Eq. 6}$$

where $E_{CW}^{cl+cap1}(n)$ and $E_{CW}^{cl+cap1}(n = 0)$ are the energies of the $n$th excited and ground electronic states of the $V_{emb}$-embedded capped cluster within CW, respectively. Martirez and Carter demonstrated the utility of capped emb-CW to calculate the metal-to-ligand charge-transfer spectrum of a ruthenium aqua bipyridine complex attached to a small titanium oxide cluster[50] meant to model a piece of a dye-sensitized solar cell, within complete-active-space self-consistent field (CASSCF)[27,28] with second order perturbation theory (CASPT2).[51-53]

*Periodic boundary calculations.* We performed all Kohn-Sham (KS) DFT calculations with periodic boundary conditions (PBCs) using the Vienna Ab-Initio Simulation Package (VASP) version 6.2.[54,55] We used a kinetic energy cut-off of 660 eV for the planewave basis set in conjunction with the projector augmented-wave (PAW) method.[56] We used the Gaussian smearing method with 0.01 eV width for the Brillouin zone (BZ) integration along with Γ-centered finite *k*-point sampling (*vide infra*), unless otherwise specified. We start with an eight-atom diamond ($C_8$) supercell (**Fig. 2**), which we structurally optimized within DFT using the recently de-orbitalized re-regularized strongly constrained appropriately normed ($r^2$-SCAN-L)[57,58] meta generalized gradient approximation (meta-GGA) exchange-correlation (XC) functional with a *k*-point mesh of (6×6×6). DFT- $r^2$-SCAN-L, in comparison to the original SCAN functional, is, by construction, faster to execute, reproduces the accuracy of the original with respect to bulk properties of solids, and improves prediction of magnetization of ferromagnetic metals.[57] We used a real-space fine fast Fourier transform (FFT) grid of 120×120×120 (corresponding to fine grid spacing of ~0.060 Å). DFT- $r^2$-SCAN-L showed exceptional lattice constant prediction compared to experiment (**Table 1**). However, as with most nonhybrid, local, semi-local, and even meta-GGA XC functionals, DFT- $r^2$-SCANL severely underpredicts the fundamental gap ($E_g$) by more than 1 eV when using the eigenenergies of the frontier band states (**Table 1**).

**Table 1**. Predicted lattice constant and indirect band gap[a] for a pristine diamond ($C_8$ supercell) from different DFT approximations compared to the experimental structure and fundamental band gap.

| Method | Cubic *a* (Å) | $E_g$ (eV) |
|---|---|---|
| Experiment | $3.56712 \pm 0.00005$[b] | $5.480 \pm 0.004$[c] |
| DFT-$r^2$-SCAN-L | 3.5677 | $4.22 \pm 0.02$ |
| DFT-HSE06 | - | $5.30 \pm 0.02$ |
| $G_0W_0$ @DFT-PBE | - | $5.50 \pm 0.02$[d] |

[a]Difference between the energies of the conduction band minimum and valence band maximum;
[b]X-ray diffraction at room temperature, ref. [59]; [c]photoemission-inverse photoemission, ref. [11]
[d]formally corresponds to the fundamental gap

Using the DFT- $r^2$-SCAN-L-optimized structure, we recalculated the electronic structure using the hybrid-DFT Heyd-Scuseria-Ernzerhof XC functional with the 2006 parameterization (HSE06).[60,61] We also performed a quasi-particle calculation via a low-scaling implementation[62] of single-shot GW[63-65] ($G_0W_0$) using the one-electron orbitals from DFT with Perdew-Burke-Ernzerhof GGA XC functional[66] ($G_0W_0$ @DFT-PBE) – see **Supplementary Methods** in the Supplementary Information (SI) for details. These two higher levels of theory predict $E_g$s in much better agreement with the experiment (**Table 1**). **Fig. S1** shows the densities of states for diamond from all three levels of theory that we calculated using a Γ-point-centered (10×10×10) *k*-point mesh and the linear tetrahedron method with Blöchl corrections for the BZ integration.[67] Based on these results, hereafter, we use DFT- $r^2$-SCAN-L for *atomic* structure optimizations and subsequently use DFT-HSE06 to calculate the ground-state *electronic* structures. Despite higher computational cost, we deemed DFT-HSE06 to be a much more reliable baseline theory than DFT-$r^2$-SCAN-L to perform $V_{emb}$ optimization, based on the accuracy of their $E_g$ predictions (**Table 1**). The optimized $V_{emb}$ is subsequently used in capped emb-CW theory (*vide infra*).

To simulate a periodic negatively charged N-doped diamond with a vacancy, we first constructed 64-, 128-, and 160-atom diamond supercells from diamond's unit cells. Specifically, we construct the 64-atom supercell by doubling all three lattice vectors of the eight-atom conventional cubic unit cell of diamond, i.e., a (2×2×2) conventional cubic supercell. To construct the 128- and 160-atom supercells, we multiply the lattice vectors of the two-atom primitive unit cell (face-centered cubic) of diamond by (4×4×4) and (4×4×5), respectively. We then removed one C atom and replaced an adjacent C with N (**Fig. 2**). We added an electron to the total electron count without imposing constraint on its location and performed spin-polarized KS DFT. We use a uniform positive background charge to compensate for the negatively charged defect and thus yield an over-all neutral model with a converged (electrostatic) Ewald sum. We expect the larger supercells to be most reliable in terms of reducing errors associated with the uniform background charge.[68-70] We optimized these charged defective structures without imposing symmetry within DFT-$r^2$-SCANL using a Γ-point-centered (3×3×3) $k$-point mesh. For the 128- and 160-atom diamond supercells, we used a real-space fine FFT grid of 180×180×180 and 180×180×216, respectively (maintaining a fine grid spacing of ~0.06 Å, *vide supra*).

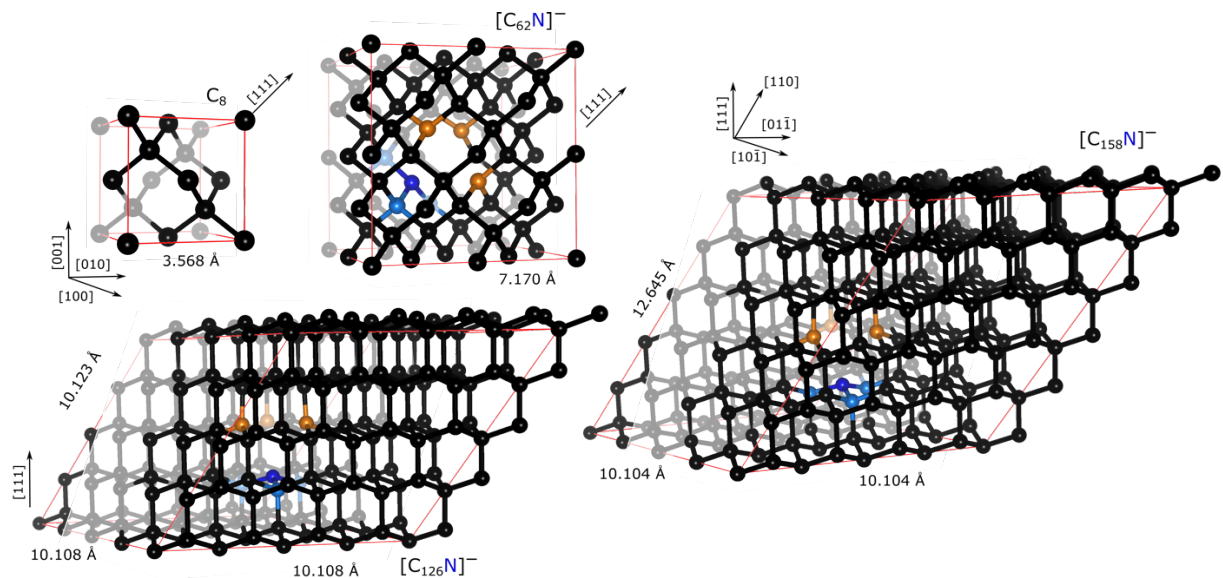

**Fig. 2**. **Bulk atomic structures of diamond** – flawless cubic ($C_8$) and with $N_CV_C^-$ defect: $C_{62}N^-$, $C_{126}N^-$, and $C_{158}N^-$ (as labelled), optimized within DFT-$r^2$-SCAN-L. The dimensions of the supercell are given (if only one is given, then all sides are equal). N – dark blue; C – black. For clarity, the three C atoms adjacent to $V_C$ are orange whereas the atoms bound to N are light blue. The red lines mark the supercell edges. The lattice vectors according to the primitive cubic cell are shown. **Table S1** summarizes the optimized cell parameters.

Upon electronic and structural relaxation, the excess electron localizes at the defect site, specifically at the three three-fold coordinated C atoms adjacent to the $V_C$ (C[3c]) as expected (**Fig. S2**). Note that a unform background charge does not correspond to an applied electric field and is added as a spatially constant potential in the electronic Hamiltonian. The Hamiltonian with and without such uniform potential will therefore yield the same electronic eigenstates. We will later come back to this property. However, despite the convergence of the Ewald sum when using a uniform compensating background charge, the Coulomb interactions between the periodic images of the charged defect converge slowly with system size: monopole-monopole (M-M) decays as $1/L$ ($L$ is distance), whereas dipole-dipole and quadruple-monopole interactions decays more quickly as $1/L^3$.[68] The long-range M-M interactions render the total energy of charge defects also slowly convergent with system size. Although correction schemes to the total energy of charged supercells with compensating uniform background charge are available,[68,71] we avoid altogether the use of the total energy calculated from the PBC calculations and simply used PBC to obtain an optimized $V_{emb}(\mathbf{r})$.

To optimize $V_{emb}(\mathbf{r})$, we select a suitable cluster (and environment) that we carve out from the periodic defective bulk diamond (see **Fig. 2** for the bulk structures). We capped the edges of the resulting $C_xN$ cluster and $C_y$ environment using fluorine (F), oxygen (O), and boron (B) atoms for when respectively monovalent, divalent, and trivalent capping is needed (more details in the **Results and Discussion** section). We used a modified VASP v. 6.4.2 to perform $V_{emb}(\mathbf{r})$ optimization (*vide supra*),[72] which utilizes a DFET implementation within the PAW method with its frozen-core approximation by Yu, Libisch, and Carter.[47] In $V_{emb}(\mathbf{r})$ optimization for the negatively charged defect, the periodically repeated capped cluster contains the excess charge and is solved for as a spin triplet within the spin-unrestricted (broken spin symmetry) formalism - the same as the full periodic bulk system. The auxiliary fragment and capped environment are charge-neutral and solved for as closed-shell within the spin-restricted formalism. The periodic charged capped embedded cluster (with compensating uniform background charge) is subject to the same slowly converging ($1/L$) monopole-monopole interaction as the full system because they share the same supercell size and excess charge distribution: the excess charge is localized around the defect (**Fig. S2**) and at convergence of $V_{emb}(\mathbf{r})$, the regional matching of the electron density between the embedded cluster and full system, and by extension, the excess charge distribution, is ensured. Without loss of generality, the same equality can be said about the local defect-induced dipole-dipole and monopole-quadrupole interactions when the capped cluster is chosen such that it preserves the symmetry of the defect in the full system (which we demonstrated to be true, see **Note S1** and **Table S2**). Therefore, the optimized $V_{emb}(\mathbf{r})$ enacts the correct local dielectric response of the cluster to the environment (and environment to the cluster) not captured by the capping elements, free of the slowly convergent Coulomb interactions.

*Cluster calculations*. We performed all bare and embedded cluster calculations using Molpro version 2024.1.1.[73,74] We employed atom-centered Gaussian-type orbital (GTO) basis sets. GTO bases include Dunning[75] augmented correlation-consistent polarizable double ζ (aug-cc-pvdz or apvdz), which we benchmarked against the aug-cc polarizable triple ζ (aug-cc-pvtz or apvtz) basis set. **Table S3** summarizes the basis sets we used throughout the study. We concurrently used the complementary auxiliary bases to evaluate the two-electron integrals (so-called density fitting). Accordingly, we transformed the optimized $V_{emb}$ from (real-space) uniform Cartesian grid (as produced within VASP) to the GTO bases space via the embedding integral generator code (EmbeddingIntegralGenerator).[76] We then add the $V_{emb}$ in GTO to the one-electron Hamiltonian. Because the cluster is nonperiodic, a neutralizing background charge is not needed to converge the electrostatic energy, furthermore the distance-dependent Coulomb interaction between a charge defect and its periodic images are non-existent (*vide supra*). Furthermore, with or without the uniform charge-screening potential, the eigenstates will be the same, wherein the former all the eigenvalues (and hence the total energy) are simply shifted by a constant relative to the latter. Therefore, in calculating the excitation energies, the difference between the ground and excited states for the same number of electrons (**Eq. 6**) will lead to the perfect cancellation of the energy shift. Therefore, even though the cluster defect is charged, we choose not to apply the same uniform background charge (potential) as we have used in cluster PBC (where it is necessary).

We performed active space optimization using the molecular orbital theory as guide to determine the appropriate correlation space (*vide infra*). We performed state-averaged CASSCF (SA-CASSCF), averaging over three states, to calculate for the ground and lowest valence excited states. We did not impose any symmetry. Three states are needed due to the double degeneracy of the first excited state of the spin triplet and the ground state of the spin singlet state. Because CASSCF is only able to capture static correlation within the limited correlation space (active space) size, we captured the dynamic correlation via an MRPT2 method, specifically, NEVPT2.[29] We also performed CASPT2[51-53] in some cases as an alternative method. Despite having no intruder states, we found an ionization potential-electron affinity (IP-EA) shift[77] of 0.25 ha is required in CASPT2 to avoid overprediction of the excitation energies (as is found in the literature). We concluded NEVPT2 to be superior to CASPT2 because it does not require such a semi-empirical tuning parameter (*vide infra*). We therefore used NEVPT2 in the entirety of the paper.

**Results and Discussion**

*Bulk and defect electronic structure within DFT.* **Fig. 3A** shows the calculated DFT-HSE06 densities of states (DOS) for a flawless diamond using a $C_{64}$ supercell. **Fig. 3B** shows the corresponding DOS for a diamond with $N_CV_C^-$ defect for a $C_{62}N^-$ supercell (**Fig. 2**). **Fig. S1** shows the DFT-$r^2$-SCAN-L, DFT-HSE06, and $G_0W_0$ @DFT-PBE DOS for the flawless $C_8$ diamond structure for comparison. For pure diamond, the DOS plot is centered, i.e., zeroed, midway between the valence band (VB) and conduction band (CB) edges – the band gap center, which is also the Fermi level ($E_F$). The DOS of the pristine diamond clearly show the expected band gap, closely matching the experimental fundamental gap (*vide supra*). The pristine diamond frontier band states are simply the C-C bonding (VB) and anti-bonding states (CB). For the defective structure DOS (**Fig. 3B**), we match the position of the CB minimum (CBM) with the CBM of the pure diamond of the same supercell size (**Fig. 3A**). Furthermore, for the defective diamond, the spin majority (up) and minority (down) states are separately plotted. The $E_F$ shifts from the band gap center to the highest occupied defect state in the spin-up channel (**Fig. 3B**). The DOS profile appears to be only weakly sensitive to supercell size from $C_{62}N^-$ to $C_{158}N^-$ supercell - we performed this analysis only within DFT- $r^2$-SCAN-L (**Fig. S3**) because of the tremendous computational cost of a dense $k$-point mesh that would be required to generate a corresponding DFT-HSE06 DOS for the $C_{126}N^-$ and $C_{158}N^-$ supercells.

**Fig. 3B** also shows the projected DOS (pDOS) of the third-nearest-neighbor C atoms of the defects and beyond, which exhibit an almost pristine-diamond-like profile. To further resolve atomic contributions in the frontier states, **Fig. 3C** plots the pDOS of the C atoms adjacent to $V_C$ (three-fold coordinated C or C[3c], orange) and their nearest neighbors (teal), C atoms next to the substitutional $N_C$ (C[nN], light blue), and $N_C$ (dark blue). In the same plot, the pDOS of the second-nearest neighbors of the defects (C[3c] and N) are also shown (grey). **Fig. 3D** illustrates the locations of the different types of C atoms relative to C[3c] and $N_C$, showing atoms in the same color-coding scheme as their pDOS. **Fig. 3C** shows that the diamond with $N_CV_C^-$ defect exhibits pristine-like pDOS for the C atoms away from the defect (including the second-nearest-neighbor C atoms). The highest occupied state of the N dopant (corresponding its lone pair) is near the VB edge. The mid-gap states are defect states associated with the "dangling bonds" of the three C[3c]s with some contributions from a N 2$p$ state. The lower energy mid-gap state is doubly occupied (both up and down spin states are occupied) – an $a_1$ state (*vide infra*), which has contributions from C[3c]s and N. The higher energy doubly degenerate gap states are singly occupied (their corresponding spin down states are above $E_F$) - $e$ states (*vide infra*), which has contributions exclusively from C[3c]s. The two unpaired electrons localized at three C[3c]s give rise to the $V_C$-centered triplet spin multiplicity (**Fig. S2**). The approximate spin triplet defect excitation energy is the difference in the eigenenergies of the $e$ and $a_1$ states in the spin down manifold with the $e$ orbitals being empty in the ground state for this spin channel, which is 2.38 eV (**Fig. 3**). The defect excitation energy depends weakly, although non-negligibly, on supercell size from $C_{62}N^-$ to $C_{158}N^-$, with the value decreasing by 0.08 eV from the smallest to the biggest supercell within DFT- $r^2$-SCAN-L (**Fig. S3**). We will show later that excitation energies from our embedded capped cluster method are even less sensitive to supercell size.

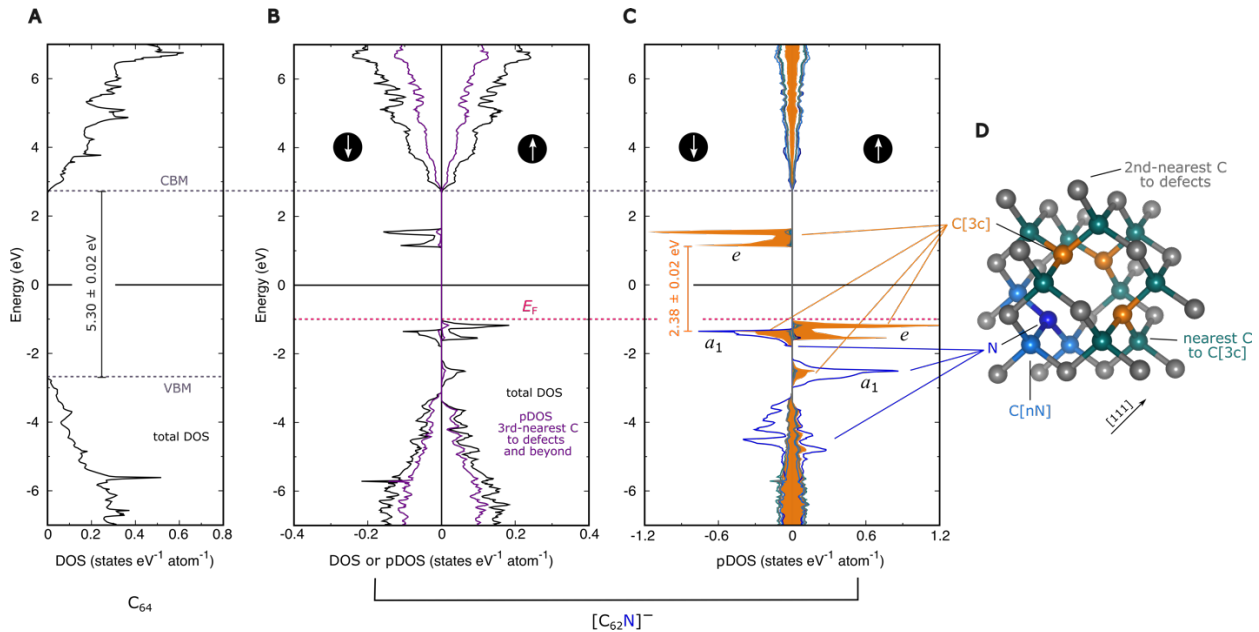

**Fig. 3**. **Bulk electronic DOS of diamond ($C_{64}$) and diamond with a $N_CV_C^-$ defect ($C_{62}N^-$) from DFT-HSE06.** The DOS are normalized by the total number of atoms in the supercell. The pDOS are normalized over the number of atoms whose atomic contributions are summed. The DOS of the defective structure (**B** and **C**) shows separately the spin down (left) and up (right) manifolds. **B** also shows the pDOS for the C atoms far from the defect (as labelled). **C** shows the contributions of the defect atoms along with their nearest neighbors (as shown in **D**). **D** shows a partial structure of the supercell annotating the position of the defect atoms and their neighbors. Optimized atomic structures from DFT-$r^2$-SCAN-L were used. Note the difference in DOS scale between plots.

This (Kohn-Sham DFT) orbital-energy-based method of approximating the defect excitation energies in solids is widely adopted in the literature, with mixed success, and is naturally dependent on the XC functional used.[78-80] Excitation energies based on one-electron orbital energies are at best an approximation of the energy difference between adding an electron and taking away an electron from and to the vacuum, *i.e.*, ionization energy minus electron affinity (IP-EA) known as the fundamental gap.[81] The IP approximation is based on Koopmans' theorem that the energy of the highest occupied one-particle (Hartree-Fock) orbital of a molecule is its negative IP,[82,83] which was later extended to include the approximation that the energy of the lowest energy empty one-particle orbital is the molecule's EA (extended Koopmans' theorem) – both at the limit of no orbital relaxation upon electron addition or removal.[84] These approximations were later generalized to one-particle orbitals from Kohn-Sham density functional theory for molecules and solids.[85-87] A local defect-centered excitation, on the other hand, corresponds to the energy needed to form an exciton (bound electron and hole) – optical gap.[81] The optical (excitonic) gap should be lower than the fundamental gap by the exciton binding energy.[81] The experimental vertical excitation energy (VEE) of the $N_CV_C^-$ defect was reported to be around 2.18 eV – lattice relaxation in the excited state preceding an electronic relaxation (emission) lowers the energy of emitted light, corresponding to the sharp zero phonon line (ZPL) in the fluorescence spectrum: 1.95 eV for $N_CV_C^-$.[9,88] The electronic structures predicted by DFT-HSE06 closely match the experimental and quasi-particle method ($G_0W_0$) band gap of pure diamond (5.30 vs. 5.50 eV; **Table 1**, **Figs. S1** and **3A**) and provides a reasonable upper bound to the VEE of the spin triplet state of the defect (2.38 eV, **Fig. 3C**) when using DFT- $r^2$-SCAN-L-optimized atomic structures. We wish to evaluate the VEE of this defect within a more accurate and formally excited-state theory that explicitly solves for the excited-state wavefunction and incorporates the multiconfigurational nature of the spin singlet state of the defect. We next describe the partitioning scheme we used and the embedded capped cluster calculations it enabled.

*System partitioning and $V_{emb}$ optimization*. The charge and spin density difference plots in **Fig. S2** and pDOS in **Fig. 3C** offer guidance on how to construct a suitable cluster. **Fig. S2** shows that the excess charge and spin are highly localized at the three C[3c]s. Further, the atomic pDOS (**Fig. 3C**) illustrates the degree of the contributions to the defect states (labelled $a_1$ and $e$) of the C atoms surrounding the defect atoms (C[3c]s and N) and suggests that the C atoms nearest to the defect atoms (teal and light blue C atoms in **Fig. 3D**) adequately capture the spatial extent of the electronic influence of the defect. Therefore, starting from the full $C_{62}N^-$ supercell, we carved out a $C_{15}N$ cluster where the $C_{15}$ includes the three C[3c]s, three C[nN], and nine nearest neighbors of C[3c]s. We designate the remaining C atoms ($C_{47}$) as the environment fragment (**Fig. 4**, left panel). The $C_{15}N$ yields 12 coordinatively unsaturated C atoms at the edges, with each having to share a missing common neighbor *individually* with two other edge C atoms. We replaced the unshared missing C atoms with F atoms (terminal or monovalent capping) – a total of 12, and the doubly shared missing C atoms with O atoms (bridge or divalent capping) – also a total of 12, to yield a $C_{15}NF_{12}O_{12}$ capped cluster (**Fig. 4**, left panel). On the other hand, the environment fragment contains 24 edge C atoms that are missing a C neighbor, however each of them shares a missing neighbor *jointly* with two other edge C atoms. We replaced the triply shared missing C atoms with B atoms (facial or trivalent capping) – a total of 12, to yield a $C_{47}B_{12}$ capped environment (**Fig. 4**, left panel). This capping scheme results in fragments that are coordinatively saturated at the fragmentation points. The auxiliary fragment composed of combined capping elements from both the cluster and environment ($F_{12}O_{12}B_{12}$) also results in coordinatively saturated closed-shell hypothetical molecular fragments (**Fig. 4**, left panel), which is a consequence of valence complementarity principle we abide by in deciding what elements to cap the cluster and environment fragments. Specifically, the auxiliary fragment contains four closed-shell $F_3O_3B_3$ units, where each B atom bonds with two O and one F atom.

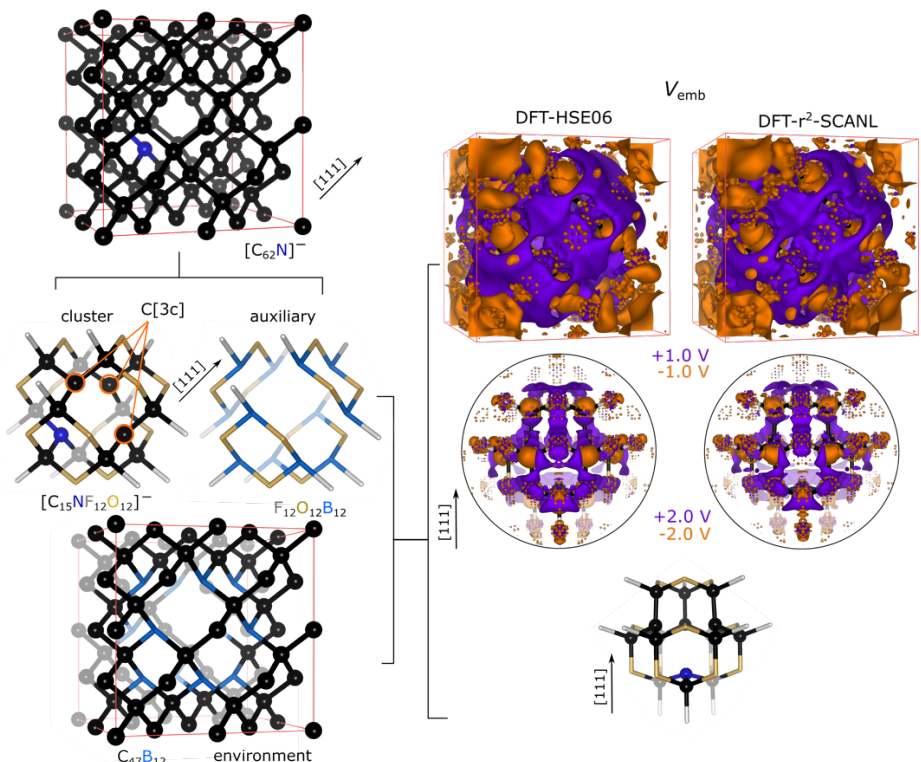

**Fig. 4**. **Partitioning and embedding schemes**. Left panels, partitioning and capping scheme and optimized $V_{emb}$ for the $C_{62}N^-$ periodic supercell. Cluster, environment, and auxiliary fragments' structures shown as labelled. Right panels, optimized $V_{emb}$ within DFT-HSE06 and DFT-r$^2$-SCANL as labelled at two isosurface levels. $V_{emb}$s feature repulsive (purple) and attractive (orange) potentials.

We optimize the $V_{emb}(\mathbf{r})$ (see **Theory and Methods**) following the recipe of the capped-DFET scheme by Martirez and Carter[26] using $C_{15}NF_{12}O_{12}$ with an excess electron ($[C_{15}NF_{12}O_{12}]^-$), $C_{47}B_{12}$, and $F_{12}O_{12}B_{12}$ to reconstruct the charge density of $C_{62}N^-$ within DFT-HSE06, as well as DFT-r$^2$-SCAN-L for comparison. The optimized $V_{emb}(\mathbf{r})$s shown in **Fig. 4**, right panel features a repulsive potential shell surrounding the defect cluster with a complementary attractive potential that limits the spillover of the electron density of the environment into the empty defect region. Conversely, the positive potential surrounding the defect cluster partially delocalizes the electron density from the periphery of the cluster into the environment region. At higher absolute potentials, the potential repolarizes the C–O and C–F bonds to offset the electron pull of O and F (notice the attractive potential at the boundary C and positive potential surrounding O). The $V_{emb}(\mathbf{r})$s from the two XC functionals share the same features (**Fig. S4** shows their difference). Using these optimized $V_{emb}$, we performed MRPT2 calculations for the $[C_{15}NF_{12}O_{12}]^-$ cluster to calculate excitation energies for both spin triplet and singlet. Note that we use the same $V_{emb}(\mathbf{r})$ optimized for the ground-state spin triplet for the embedded capped cluster spin singlet to retain the same Hamiltonian for both spins – we use after all also the same spin-triplet optimized structure. The use of the same $V_{emb}(\mathbf{r})$ enables us to directly compare energies of the two spins without the risk of artifacts that may originate from two different $V_{emb}$s. We thus assume that the dielectric screening of the environment in response to defect transitions (electronic excitation and spin crossover) will not change significantly as long as the atomic structure remains the same.

*Defect electronic structures within emb-capped-CW.* To initiate an MRPT2 calculation, we begin with a series of CASSCF calculations to determine the dominant electronic configurations for both ground and excited states for the two spin states, as well as to construct a set of self-consistently optimized many-body wavefunctions that will be the basis for the multi-configurational second-order perturbation energy correction. CASSCF requires one to specify an active space (AS) size, *i.e.*, number of valence electrons and a set of orbitals to generate a series of electronic configurations to define a set of determinants to be used to expand the wavefunction(s). Because the (canonical) orbitals are self-consistently optimized along with the contributions of each electronic configuration to the overall wavefunction (configuration interaction, CI), CASSCF captures self-consistently the (static) electronic correlation that is important to describe degenerate/near degenerate states as well as to robustly represent excited states. Because we are interested in calculating the electronic transitions involving the $N_CV_C^-$ defect, we consult the molecular orbital (MO) diagram resulting from the interaction of N atom with six C[3c]s at two neighboring $V_C$s (**Fig. 5A**). As the N atom fills one of the vacancies, N uses its valence orbitals (2*s*2*p*) to mix with three (2*sp*$^3$-like) dangling orbitals of the C[3c]s, yielding three N-C bonding (σ) and antibonding (σ$^*$) covalent single bonds, and a

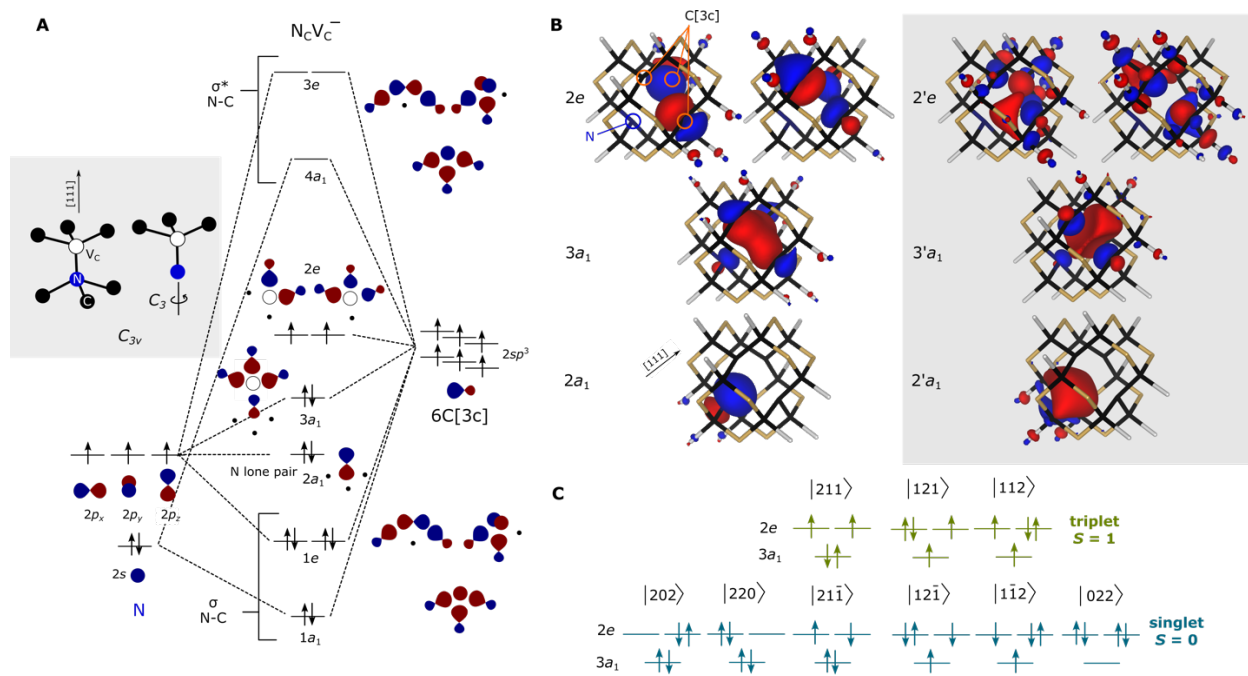

**Fig. 5**. **Defect MO, calculated orbitals, and dominant electronic configurations**. **A**, grey box: simplification of the structure of the defect site emphasizing the $C_{3v}$ symmetry. Corresponding molecular orbital diagram of the $NV_CC_6$ defect. The defect's $C_{3v}$ symmetry dictates orbital and degeneracy of the electronic states. The valence $3a_1$ and $2e$ orbitals originate from the non-bonding $sp^3$ states of the three carbon atoms at the vicinity of the vacancy. **B**, natural frontier orbitals $a_1$ and doubly degenerate $e$ calculated from the state-averaged density matrix of the three-state triplet embedded SA-CASSCF with CAS(6e,8o) from apvdz basis set. Isosurface value of 0.06 (e/bohr$^3$)$^{1/2}$. **C**, electronic configurations for the spin triplet and singlet states showing only the occupation of the valence $3a_1$ and $2e$ orbitals with spin-up and spin-down electrons (up and down arrows). In the bra-ket notation, $2, 1, \bar{1}$, or $0$ mean that an orbital is either doubly, spin-up singly, spin-down singly, or unoccupied, respectively. Note that these electronic configurations are not necessarily degenerate and in fact are not.

N-centered lone pair. Because the $N_CV_C^-$ defect exhibits a local symmetry of $C_{3v}$, we label non-degenerate σ and σ* as $1a_1$ and $4a_1$ (invariant with respect to $C_3$ rotation and reflection), respectively, whereas we label the doubly degenerate σ and σ* as $1e$ and $3e$, respectively. The N lone pair, also possess an $a_1$ symmetry, and thus, labelled as $2a_1$. The remaining three $2sp^3$-like C[3c] orbitals linearly mix where all three are in phase ($3a_1$) and a doubly degenerate state where two $2sp^3$ orbitals are out of phase ($2e$). For the ground state $N_CV_C^-$, $3a_1$ is doubly occupied, whereas $2e$ are each singly occupied, hence the spin triplet (two unpaired electron) magnetization. If one considers all the aforementioned orbitals (and electrons) in the AS, the AS size will be 12 electrons in 10 orbitals or (12e,10o). Furthermore, for the excited states, higher energy virtual orbitals may correlate with the lone-pair $2a_1$ and vacancy-centered $3a_1$ and $2e$ orbitals non-trivially. If these four orbitals are also to be included, labelled $2'a_1$, $3'a_1$, and $2'e$, then the full AS size would be (12e,14o) – a gargantuan task even with the reduced size of the system enabled by embedding. We therefore systematically increase the AS in the CASSCF calculations to determine the sensitivity of the calculated excitation energies: (4e,6o) with 6o = ($3a_1$, $2e$, $2'e$, $3'a_1$); (6e,8o) with 8o = ($2a_1$, $3a_1$, $2e$, $2'e$, $2'a_1$, $3'a_1$); (10e,12o) with 12o = ($1e$, $2a_1$, $3a_1$, $2e$, $2'e$, $2'a_1$, $3'a_1$, $3e$) and the full (12e,14o). We found (6e,8o) to be the optimal AS that balances precision and computational cost (*vide infra*). We calculated three states for each spin (based on local $C_{3v}$ symmetry and frontier orbital symmetry and occupation, **Fig. 5A**): for $S = 1$, the ground state is labelled $^3A_2$ and the first excited state is a doubly degenerate $^3E$; for $S = 0$, the doubly degenerate ground state is labelled $^1E$ and the first excited state is $^1A_1$.

We confirm via continuous symmetry measure[89] implemented within the VASPkit code[90] that our $[C_{15}NF_{12}O_{12}]^-$ cluster possesses a $C_{3v}$ symmetry, i.e., invariant toward $E$, two $C_3$, and three $\sigma_v$ symmetry operations.[89] **Fig. 5B** shows the optimized state-averaged natural orbitals for a (6e,8o) AS from a three-state $S = 1$ CASSCF calculation using the apvdz basis set. **Fig. S5** shows the corresponding state-averaged natural orbitals for a (6e,8o) AS from a three-state $S = 0$ CASSCF calculation, which resemble those of $S = 1$. Despite not explicitly imposing symmetry in the calculations, the natural orbitals in **Fig. 5B** exhibit the expected symmetry of the ($2a_1$, $3a_1$, $2e$) orbitals (linear combination of vacancy 2p states) in **Fig. 5A**, although the natural orbitals go beyond the single-particle picture provided in MO diagrams. The virtual ($2'a_1$, $3'a_1$, $2'e$) as expected exhibit the same approximate symmetry of the ($2a_1$, $3a_1$, $2e$) orbitals but have more nodes (linear combination of 3p-like orbitals). **Figs. S6** and **S7** show the state-averaged natural orbitals from three-state $S = 1$ with AS of (10e,12o) and (12e,14o) that now feature some or all the C–N $\sigma$ ($1a_1$, $1e$) and $\sigma^*$ ($4a_1$, $3e$) orbitals. For the (6e,8o) AS, all the active (natural) orbitals participate, *i.e.*, their occupation number changes relative to the dominant ground-state configuration, in at least one electronic configuration whose absolute CI coefficient ($c$) is ≥ 0.05 for at least one of the three calculated states. The dominant configurations are shown in **Fig. 5C** in the basis of the ($2a_1$, $3a_1$, $2e$) natural orbitals shown in **Figs. 5B and S4** for both $S = 1$ and 0. These configurations are used to build the ground and excited state wavefunctions and account for 93-96 % configuration state functions used in the expansion of the wavefunction (**Table 2**).

**Table 2.** CI coefficients ($c$) in the state-averaged natural orbital basis from SA-CASSCF (6e,8o) AS.[a]

| State | $c$ ($3a_1$, $2e$) | | | | | | $\sum |c|^2$ |
|---|---|---|---|---|---|---|---|
| | | | S = 1 | | | | |
| | | |$|211\rangle$|$|121\rangle$|$|112\rangle$| | |
|$^3A_2$| | | 0.978 | 0.000 | 0.000 | | 0.956 |
|$^3E$| | | 0.000 | 0.174 | 0.949 | | 0.931 |
| | | | 0.000 | 0.949 | -0.174 | | 0.931 |
| | | | S = 0 | | | | |
| |$|202\rangle$|$|220\rangle$|$|21\bar{1}\rangle$|$|12\bar{1}\rangle$|$|1\bar{1}2\rangle$|$|022\rangle$| |
|$^1E$| -0.468 | 0.468 | 0.564 | -0.104 | -0.425 | 0.000 | 0.948 |
| | 0.399 | -0.399 | 0.662 | 0.425 | -0.104 | 0.000 | 0.948 |
|$^1A_1$| 0.570 | 0.570 | 0.000 | 0.000 | 0.000 | -0.532 | 0.933 |

[a]Electronic configurations illustrated in **Fig. 5C**, SA-CASSCF natural orbitals are shown in **Fig. 5B** ($S = 1$) and **Fig. S5** ($S = 0$)

For $S = 1$, a single determinant ($|211\rangle$) accounts for 96 % of the $^3A_2$ state. A single excitation from the $3a_1$ to $2e$ orbital (some constructive and destructive linear combination of the $|121\rangle$ and $|112\rangle$ configurations) – accounting for 93 % electron correlation – yields the $^3E$ states. For $S = 0$, the ground and excited states are more truly multiconfigurational with four and three configurations, respectively, almost equally dominating the wavefunction expansion. This easily explains the failure of single-determinant methods (e.g., DFT) to describe even the ground state of the spin singlet $N_CV_C^-$. Of note, the ground $^1E$ states are two destructive combinations of the $|202\rangle$ and $|220\rangle$ configurations, whereas $^1A_1$ is the constructive combination of the two configurations. The $^1E$ states also include $|21\bar{1}\rangle$, $|12\bar{1}\rangle$, and $|1\bar{1}2\rangle$, which are spin flipped ground and excited states configurations of $S = 1$. Also, of great significance is that $^1A_1$ includes a non-trivial contribution ($|c|^2 \sim 0.28$) from a double excitation $|022\rangle$, therefore, methods that only account for single excitations (e.g., time-dependent DFT) to simulate the first excited state of the spin singlet $N_CV_C^-$ will most certainly lead to incorrect predictions.[91] Furthermore, (ignoring spin-orbit coupling) the absence of contributions from the spin-flipped excited $S = 1$ configurations in $^1A_1$ state indicate that spin crossover from $^3E$ to $^1A_1$ is not simply a spin flip relaxation.

*Defect excitation energies within emb-capped-CW.* We followed the SA-CASSCF calculations with state-specific NEVPT2 (SS-NEVPT2 or simply NEVPT2) to calculate the first vertical excitation energies (VEE) for $S = 1$ and $S = 0$, as well as the excited-state singlet-triplet splitting energy, i.e., intersystem crossing (ISC) between $^1A_1$ and $^3E$, using **Eq. 6**. Note that VEE is an upper bound of the so-called zero phonon line (ZPL) because it ignores structural relaxation following the electronic transition (fixed nuclei approximation). The first VEE for $N_CV_C^-$ are also experimentally known for both $S = 1$ and 0 along with the ZPL, the former from optical absorption spectroscopy whereas the latter from photoluminescence measurements.[9,10,88,92,93] The $^1A_1 \rightarrow {}^3E$ ISC energy has also been reported in the literature, which is approximated from ISC rates.[94-96]

**Table 3** summarizes the $^3A_2 \rightarrow {}^3E$, $^1E \rightarrow {}^1A_1$, and $^1A_1 \rightarrow {}^3E$ VEEs calculated here from different basis set sizes, AS sizes, DFT approximations to derive an optimized $V_{emb}$, and MRPT2 methods. Even without imposing symmetry, the two highest energy spin triplet states are (numerically) degenerate (within 0.36 meV of each other), thus, aptly assigned as $^3E$. Likewise, the two lowest-energy spin singlet states are degenerate (difference of 0.23 meV), thus, aptly assigned as $^1E$. The same table shows the best available experimental values along with values from the computational literature that went beyond DFT eigenvalue differences in calculating excitation energies. We first focus on our emb-NEVPT2 calculations along with using the DFT-HSE06-derived $V_{emb}$. For the same basis set size, the $^3A_2 \rightarrow {}^3E$ VEE decreases by as much as 0.07 eV from an AS of (4e,6o) to (6e,8o), which demonstrates the importance of including the N lone pair ($2a_1$) and its correlating orbital in the wavefunction expansion to accurately describe the excited state. The spatial reach of the lone pair into the vacancy site and similarity of its energy to the vacancy states can explain its importance. Indeed, **Fig 5B** shows that the $3a_1$ state exhibits some N $2p$ character and DFT-HSE06-calculated pDOS (**Fig. 3C**) illustrates the contribution of N to the defect $a_1$ state. Expanding the AS to (10e,12o) only lowers the $^3A_2 \rightarrow {}^3E$ VEE by 0.01 to 0.02 eV, indicating good convergence for (6e,8o) AS. While it is possible to perform CASSCF calculations with an AS of (12e,14o), the memory requirements made it impossible to perform the subsequent MRPT2 calculation. The $^1E \rightarrow {}^1A_1$ VEE is less sensitive to the AS size, whereas $^1A_1 \rightarrow {}^3E$ ISC energy exhibits the same sensitivity as the triplet excitation energy presumably due to the sensitivity of the relative energy of the $^3E$ state with the AS size. The effect of the basis set size going from apvdz to apvtz (additional $s,p,d$ diffuse and polarization, *i.e.*, two $f$, functions are included, **Table S3**) for C and N (with the capping elements F and O remaining apvdz) led to ~ 0.01 eV (within the round-off error) reduction in VEE for the same AS.

**Table 3.** Calculated vertical excitation energies (VEE) for low-lying transitions in N$_C$V$_C^-$ center from emb-MRPT2 using an [C$_{15}$NF$_{12}$O$_{12}$]$^-$ capped cluster carved from C$_{62}$N$^-$ periodic supercell.

| | | | | Calculated VEE [eV] | | | |
|---|---|---|---|---|---|---|---|
| | | | | $V_{emb}$ (DFT) | | | |
| | | | | DFT-HSE06 | | DFT-r$^2$SCANL | **Recent literature** *(beyond DFT methods)* |
| Exc. | Exp. VEE$^a$ (ZPL)$^b$ | Basis set | AS size | Emb-NEVPT2 | Emb-CASPT2w/ IPEA shift | Emb-NEVPT2 | \|error\| ≤ 0.10 eV<br>\|error\| ≤ 0.15 eV<br>\|error\| > 0.15 eV |
| $^3A_2 \rightarrow {}^3E$ | **2.18**$^c$ (1.95)$^d$ | apvdz | (4e,6o) | 2.32 | 2.36 | 2.35 | 2.32 *GW*+BSE$^i$<br>2.05 CI cRPA$^j$<br>2.00 QET-beyond-RPA$^k$<br>2.31 DMET-NEVPT2$^l$<br>2.15 QDET EDC@$G_0W_0^m$<br>1.98 DMFET-FCIQMC$^n$<br>2.14 cluster CASSCF$^o$ |
| | | | (6e,8o) | 2.25 | 2.31 | 2.29 | |
| | | | (10e,12o) | 2.23 | 2.30 | 2.27 | |
| | | apvtz+ apvdz$^h$ | (4e,6o) | 2.30 | - | - | |
| | | | (6e,8o) | 2.24 | - | - | |
| $^1E \rightarrow {}^1A_1$ | **1.26**$^e$ (1.19)$^f$ | apvdz | (4e, 6o) | 1.28 | 1.25 | 1.27 | 0.59 *GW*+BSE$^i$<br>0.89 CI cRPA$^j$<br>1.20 QET-beyond-RPA$^k$<br>1.02 DMET-NEVPT2$^l$<br>0.81 QDET EDC@$G_0W_0^m$<br>0.99 DMFET-FCIQMC$^n$<br>1.35 cluster CASSCF$^o$ |
| | | | (6e,8o) | 1.28 | 1.26 | 1.28 | |
| | | | (10e,12o) | 1.29 | 1.27 | 1.29 | |
| | | apvtz+ apvdz$^h$ | (6e,8o) | 1.27 | - | - | |
| $^1A_1 \rightarrow {}^3E$ | (0.32-0.41)$^g$ | apvdz | (4e, 6o) | 0.43 | 0.38 | 0.47 | - *GW*+BSE$^i$<br>0.69 CI cRPA$^j$<br>0.24 QET-beyond-RPA$^k$<br>0.80 DMET-NEVPT2$^l$<br>0.88 QDET EDC@$G_0W_0^m$<br>0.41 DMFET-FCIQMC$^n$<br>0.54 cluster CASSCF$^o$ |
| | | | (6e,8o) | 0.35 | 0.33 | 0.39 | |
| | | | (10e,12o) | 0.36 | 0.32 | 0.40 | |
| | | apvtz+ apvdz$^h$ | (6e,8o) | 0.35 | - | - | |

$^a$VEE corresponds to the excitation from the structural ground state to an electronically excited state that is in a higher vibronic state. This transition corresponds to the needed energy to excite the electrons with the nuclei frozen at the ground state
$^b$Zero phonon line (ZPL) accounts for phonon relaxation following vertical excitation when the structural minima of the lower and higher states are not the same
$^c$Approximated by ref. [88] based on the sideband maximum of the visible absorption spectra measured at liquid N$_2$ temperature after 2 MeV electron irradiation at room temperature followed by annealing from ref. [9]. Furthermore, subtracting ref. [18]'s calculated VEE-to-ZPL energy correction of -0.23 eV (derived from a 63-atom *GW* with Bethe-Salpeter equation (BSE) calculation) from the experimental ZPL of 1.95 eV also yields an approximate VEE of 2.18 eV.
$^d$Sharp peak in the visible absorption spectra measured at liquid N$_2$ temperature after 2 MeV electron irradiation at room temperature followed by annealing from Ref. [9]. Fluorescence spectra at 4 K with 532 nm excitation laser from ref. [92]
$^e$Supercontinuum IR absorption spectra at 10 K with 35 mW pump laser, 0.071 eV blue shift from ZPL absorption associated with the first phonon sideband peak of the $^1E \rightarrow {}^1A_1$ excitation from ref. [92]
$^f$IR emission after 532 nm laser excitation from ref. [10], Fluorescence IR from ref. [93]
$^g$Energy window within which $^1A_1 \rightarrow {}^3E$ ISC rate was evaluated in ref. [94,95], corrected in an erratum ref. [96]
$^h$apvtz for C and N atoms; apvdz for the capping atoms, F and O (**Table S3**)
$^i$*GW* with Bethe-Salpeter equation (BSE) for a C$_{254}$N$^-$ supercell from ref. [18]. Only ZPL is available for $^1E \rightarrow {}^1A_1$.
$^j$Configuration interaction constrained RPA (CI-cRPA) for a C$_{510}$N$^-$ supercell from ref. [19]
$^k$Defect orbital embedding via quantum embedding theory (QET) with "beyond" random phase approximation (RPA) for a C$_{214}$N$^-$ supercell ref. [20]
$^l$Density matrix embedding theory (DMET) with NEVPT2 AS(10e,9o) in restricted open-shell Hartree-Fock for a C$_{15}$N impurity cluster within C$_{214}$N$^-$ supercell from ref. [21]
$^m$Localized (defect) orbital embedding via quantum defect embedding theory (QDET) with full CI and double counting correction (EDC) from $G_0W_0$ approximation for a C$_{510}$N$^-$ supercell from ref. [22]
$^n$Density matrix functional embedding theory (DMFET) with full CI quantum Monte Carlo (FCIQMC). C$_3$H$_{12}$N$^-$ embedded fragment within C$_{42}$H$_{42}$N$^-$ cluster from ref. [23].
$^o$Non-periodic C$_{85}$H$_{76}$N$^-$ cluster with (6e,6o) AS from ref. [24].

To compare our emb-NEVPT2 results to another widely used MRPT2 method, we also performed emb-CASPT2 within the apvdz basis set. The method deviates from the experiment more than emb-NEVPT2 and required an empirical IP-EA shift of 0.25 eV to achieve an acceptable level of agreement. Due to the rather arbitrary selection for the IP-EA shift, we consider the emb-CASPT2 to be less reliable. Finally, to investigate the effect of the baseline electronic structure theory, we also optimized the $V_{emb}$ within DFT-r$^2$-SCAN-L. Although it appears that the DFT-HSE06- and DFT-r$^2$-SCAN-L-derived $V_{emb}$s are similar (**Fig. 4**, right panel, and their difference in **Fig. S4**), DFT-r$^2$-SCAN-L-derived $V_{emb}$ yields a higher $^3A_2 \rightarrow {}^3E$ VEE by only 0.04 eV than when using a DFT-HSE06-derived potential. The $^1E \rightarrow {}^1A_1$ VEE is again less sensitive to the choice of the DFT XC approximation for $V_{emb}$ optimization, whereas $^1A_1 \rightarrow {}^3E$ ISC energy exhibits the same sensitivity as $^3A_2 \rightarrow {}^3E$ VEE. The somewhat similar electron densities from DFT-r$^2$-SCAN-L and DFT-HSE06, almost agreeing to within 0.005 $e$/bohr$^3$ (**Fig. S8**), is consistent with this behavior, because $V_{emb}$ depends directly only on electron density and thus on the (occupied) one-electron orbitals, not on their orbital energies. However, care must be taken when defect states incorrectly overlap with the diamond's VB and CB states due to the more than 1 eV underprediction of diamond's band gap within DFT-r$^2$-SCAN-L (**Table 1**). Such a superfluous energy placement of the defect state(s) relative to diamond's frontier bands will likely yield a completely incorrect baseline electron density from which $V_{emb}$ will be optimized. We therefore trust more the results derived using the DFT-HSE06 $V_{emb}$.

The VEE and ISC energy predictions from emb-NEVPT2 with $V_{emb}$ from DFT-HSE06 compare very well with the experimental values, with $^1A_1 \rightarrow {}^3E$ VEE overestimated by ~ 0.05 to 0.07 eV. The $^1E \rightarrow {}^1A_1$ VEE exhibits better agreement (within 0.03 eV) whereas the predicted $^1A_1 \rightarrow {}^3E$ ISC energy is within the experimental range. At least for N$_C$V$_C^-$, the use of DFT-r$^2$-SCAN-L method appears to also lead to expected errors for the VEEs (~ 0.09 to 0.11 eV). Our calculations are much better than other excited-state-theory-based methods applied to full periodic cells in the literature, e.g., *GW* with the Bethe-Salpeter equation (*GW*+BSE)[18] and CI with constrained RPA (CI cRPA),[19] which concurrently require tremendously large supercell sizes for convergence – 255- and 511-atom supercells, respectively (**Table 3**). Our calculations also perform similarly or better than other more complex embedding schemes in the recent literature that requires an orbital localization procedure (orbital embedding), e.g., density matrix embedding theory (DMET) and quantum defect embedding theory (QDET), with random-phase approximation,[20] NEVPT2,[21] and full CI[22,23] as the CW methods, while also requiring large supercells (215- and 511-atom supercells) – **Table 3**. A recent H-capped cluster-only simulation demonstrated the need for 85-C-atom cluster ($C_{85}H_{76}N^-$) to achieve 0.2 eV convergence relative to a 33-C-atom cluster ($C_{33}H_{36}N^-$) for CASSCF excitation energies.[24] None of the alternative methods in **Table 3** exhibited the same accuracy as our prediction across the board. In the next section, we will discuss the relative insensitivity with cell and cluster sizes of our predictions in contrast to the other aforementioned methods.

*Bulk- and cluster-size dependence of calculated VEEs*. Earlier (**Theory and Methods**) we discussed the slow convergence of the Coulomb interaction of the charged defects under periodic boundary conditions even when subject to a neutralizing uniform background charge. We argued that although the periodic supercells we chose in this study would have non-vanishing defect-defect Coulomb interactions, the existence of such interactions in both the full system and the periodic embedded capped cluster would render the optimized $V_{emb}$ free of such interactions, therefore, $V_{emb}$ more closely mimics a dielectric response of the environment at the dilute defect limit. Furthermore, we subsequently calculated the excitation energies from a nonperiodic embedded capped cluster that is, by construction, free of defect-defect interactions and captures only the excitation of one defect site (instead of a site and its neighbors simultaneously as in periodic calculations). To test our assertion that $V_{emb}$ is free of slowly convergent Coulomb interactions, we optimized $V_{emb}$s for the same $[C_{15}NF_{12}O_{12}]^-$ capped cluster carved out from $C_{126}N^-$ and $C_{158}N^-$ supercells (**Fig. 2**). To also test the effect of cluster size, from the $C_{126}N^-$ supercell, we construct a bigger $[C_{21}NF_{18}O_9]^-$ cluster where we added six C atoms that are nearest neighbors of C[nN], *i.e.*, second-nearest neighbors of the dopant N. Accordingly, the additional C atoms replace six capping O atoms in $[C_{15}NF_{12}O_{12}]^-$, which required additional six capping F atoms (total of 18) and three new capping O atoms (total of nine) – structure in **Fig. 6**. **Fig. S9** shows the corresponding environment and auxiliary fragments and the optimized $V_{emb}$ for capped clusters $[C_{15}NF_{12}O_{12}]^-$ and $[C_{21}NF_{18}O_9]^-$ from the $C_{126}N^-$ supercell.

**Table 3** summarizes the calculated VEEs for the $[C_{15}NF_{12}O_{12}]^-$ and $[C_{21}NF_{18}O_9]^-$ capped clusters from the three bulk sizes with and without DFT-HSE06-derived $V_{emb}$. In the presence of $V_{emb}$, the spin triplet VEE is independent of the cluster and bulk size of origin, whereas the bare (embedding-free) case exhibits some sensitivity. Indeed, this is the hallmark of an effective embedding scheme for a localized defect. Note that although the bare capped cluster appears to compare better with the experiment (**Tables 2 and 3**), it falls short in predicting the spin-singlet VEE. The embedded capped cluster performs better for the spin singlet than the bare cluster and exhibits weaker cluster size dependence (**Table 3**). The underestimation of the excited spin singlet state in the bare cluster results in a higher $^1A_1 \to {}^3E$ ISC energy compared to the embedded capped cluster, although is less sensitive to system size. The results in **Table 3** further support (in addition to **Note S1** and **Table S2**) our postulation that our model approaches the dilute defect limit even when using an optimized $V_{emb}$ from a finite periodic model subject to slowly convergent defect-defect Coulomb interactions. Furthermore, while there is some cluster-size dependence on VEEs, they are ≤ 0.05 eV, whereas it is 0.07 for the excited-state ISC energy. **Table 3** also shows that the embedded capped cluster is overall superior to the bare capped cluster. That said, the capping procedure we introduced here alone yields acceptable levels of (in)accuracy even without embedding.

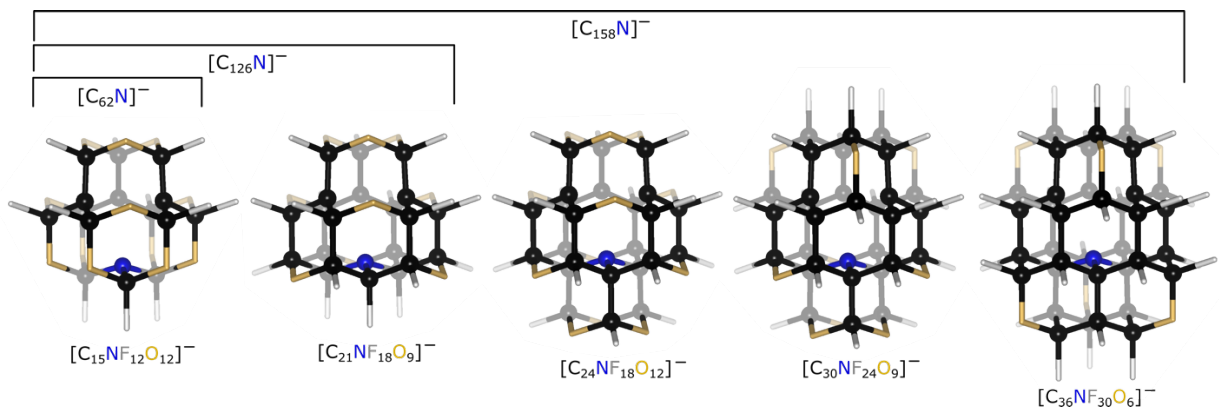

**Fig. 6**. Various clusters explored (labeled below) carved from three periodic supercells (labeled above). Atom colors match the colors of their element symbol in their labels.

**Table 3. Calculated vertical excitation energies (VEE)** for low-lying transitions in $N_CV_C^-$ center from SS-NEVPT2 (6e,8o) AS using apvdz basis set and DFT-HSE06-derived $V_{emb}$s for different bulk and cluster sizes.

| | Cluster | | Calculated VEE [eV] | |
|---|---|---|---|---|
| **Exc.** | **Bulk origin** | **Size** | $V_{emb}$ (DFT-HSE06) | **Bare** |
| $^3A_2 \to {}^3E$ | $C_{62}N^-$ | $[C_{15}NF_{12}O_{12}]^-$ | 2.25 | 2.21 |
| | $C_{126}N^-$ | $[C_{15}NF_{12}O_{12}]^-$ | 2.25 | 2.23 |
| | | $[C_{21}NF_{18}O_9]^-$ | 2.24 | 2.12 |
| | $C_{158}N^-$ | $[C_{15}NF_{12}O_{12}]^-$ | 2.25 | 2.23 |
| $^1E \to {}^1A_1$ | $C_{62}N^-$ | $[C_{15}NF_{12}O_{12}]^-$ | 1.28 | 1.18 |
| | $C_{126}N^-$ | $[C_{15}NF_{12}O_{12}]^-$ | 1.27 | 1.18 |
| | | $[C_{21}NF_{18}O_9]^-$ | 1.22 | 1.10 |
| | $C_{158}N^-$ | $[C_{15}NF_{12}O_{12}]^-$ | 1.28 | 1.18 |
| $^1A_1 \to {}^3E$ | $C_{62}N^-$ | $[C_{15}NF_{12}O_{12}]^-$ | 0.35 | 0.44 |
| | $C_{126}N^-$ | $[C_{15}NF_{12}O_{12}]^-$ | 0.37 | 0.45 |
| | | $[C_{21}NF_{18}O_9]^-$ | 0.44 | 0.46 |
| | $C_{158}N^-$ | $[C_{15}NF_{12}O_{12}]^-$ | 0.36 | 0.45 |

Other embedding methods based on orbital projections or orbital localization to partition the electrons between a region of interest and its environment (e.g., the ones listed in **Table 2**) still often require large supercells to reach convergence due to the inclusion of the slowly convergent Coulomb interaction in the full embedding Hamiltonian, which remains periodic (a large supercell also forgoes the need for *k*-point sampling, which simplifies implementation of orbital embedding for a periodic nonmetallic system). Naturally, fully periodic excited-state methods suffer from the same slowly vanishing charged defect-charged defect interactions and thus also require large supercell sizes, compounding the computational cost. Here we demonstrate that for the same cluster size, $[C_{15}NF_{12}O_{12}]^-$, the calculated VEEs and excited-state ISC energy is invariant with supercell size: $C_{62}N^-$ vs. $C_{126}N^-$ vs. $C_{158}N^-$ (greatest sensitivity is expected going from small to medium supercell sizes, which eventually plateaus for large supercells – roughly $L^{-1}$ + $L^{-3}$ dependence).[68,71] From the results above, the embedding scheme appears to be converged with respect to the capped cluster and bulk supercell sizes for the chosen model sizes. The fact that we perform the embedded CW calculation with an embedded capped cluster subject to an open boundary and an optimized $V_{emb}$ free of the periodic excess charge interactions (the reason for size convergence tests), is an advantage. **Table 3** however begs the question of whether there would be a bare capped-cluster size that will approach the embedded-capped-cluster values that is computationally tenable.

*Embedding-free cluster model convergence.* **Fig. 6** also shows three bigger capped cluster sizes in addition to $[C_{15}NF_{12}O_{12}]^-$ and $[C_{21}NF_{18}O_9]^-$, namely, $[C_{24}NF_{18}O_{12}]^-$, $[C_{30}NF_{24}O_9]^-$, and $[C_{36}NF_{30}O_6]^-$, which we generated from an even bigger $C_{158}N^-$ supercell (**Fig. 2**). Note that we maintain the $C_{3v}$ symmetry of the capped cluster even with the new atomic additions to yield bigger clusters. We systematically increase the number of C atoms in the capped clusters first by completing the first nearest neighbors of C[nN] to be all C atoms, as in $[C_{24}NF_{18}O_{12}]^-$. For $[C_{30}NF_{24}O_9]^-$, we replace the six O capping atoms of $[C_{24}NF_{18}O_{12}]^-$ with C atoms in the upper half of the cluster (vacancy side), thus recovering in the process six second-nearest C neighbors of C[3c]s. Finally, for $[C_{36}NF_{30}O_6]^-$, we replace the six O capping atoms of $[C_{30}NF_{24}O_9]^-$ with C atoms in the bottom half of the cluster (N side), thus recovering in the process six second-nearest C neighbors of C[nN]s. After these C additions, we follow the same capping principles as before: we place an F or O atom at the "missing" C sites when terminal (monovalent) or bridge (divalent) capping is required, respectively. We used these capped cluster structures to investigate the convergence of the VEEs with respect to the size of the cluster without embedding. However, the capped clusters with C atoms ≥ 24 (≥ 55 total atoms) require a smaller basis set to be computationally feasible. Capped clusters with C atoms ≥ 30 (≥ 64 total atoms) demand jointly a smaller basis set and AS to accommodate NEVPT2 memory requirements. To make possible direct comparison between different cluster sizes, we used the largest possible basis set size and AS we can do with $[C_{36}NF_{30}O_6]^-$, namely, pvdz for C and N with a smaller valence double ζ (vdz) - **Table S3**, and an AS of (4e,3o) with 3o = ($3a_1$, $2e$), which is the smallest possible correlation space to include at least ~ 93% of all CI contributions relative to the (6e,8o) AS for all of the states of interest (**Table 2**).

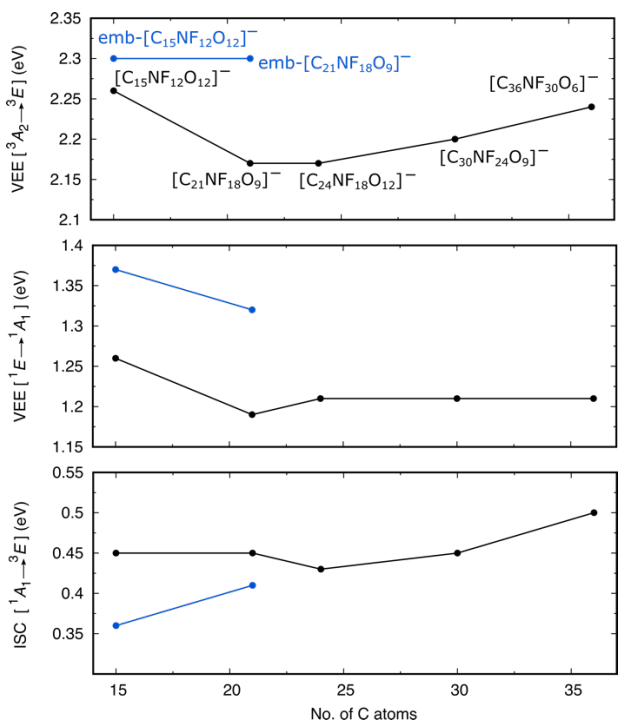

**Fig. 7**. **NEVPT2 calculated VEEs and ISC energy as a function of bare capped cluster size** (carved from a bulk $C_{158}N^-$ supercell) using pvdz+vdz basis set with (4e,3o) AS. Emb-NEVPT2 benchmarks from a bulk $C_{126}N^-$ are shown for comparison (in blue). Data summarized in **Table S4**.

**Fig. 7** plots the spin triplet and singlet VEEs along with the excited-state ISC energy as a function of the bare capped cluster size (black data points and curves). Note that due to the small basis set and AS the description of the excited state will be poorer than the ground state, and thus, the VEEs increase relative to bigger basis set and AS (**Table S4** shows the values for when using the apvdz basis set whereas **Table 3** shows values for apvdz with (6e,8o) AS for the two smallest clusters). We also calculated embedded capped cluster benchmarks for the two smallest clusters using structures and $V_{emb}$ from the $C_{126}N^-$ supercell along with the smaller basis set (pvdz+vdz) and (4e,3o) AS. **Fig. 7** also plots these embedded capped cluster benchmarks (light blue data points and curves).

The calculated spin triplet VEE from bare capped clusters dips for $[C_{21}NF_{18}O_9]^-$ (as seen in the previous section), which gradually increases back to almost the value for $[C_{15}NF_{12}O_{12}]^-$ at $[C_{36}NF_{30}O_6]^-$. Based on the trend in **Fig. 7**, the bare cluster may reach the embedded $[C_{15}NF_{12}O_{12}]^-$ but for a cluster size too large to perform an MRPT2 calculation (or any other CW method) with the proper basis set and adequate correlation space size, which thus highlights the advantage of DFET embedding. As for the spin singlet VEE, the bare cluster reaches an apparent convergence at $[C_{24}NF_{18}O_{12}]^-$ but remains below embedded $[C_{21}NF_{18}O_9]^-$ by 0.07 eV. The ISC energy between the excited-state spin triplet and singlet is less sensitive to system size, likely due to error cancellation between the two excited spin states for a given cluster size. Can a larger cluster without embedding achieve the same quality of predictions as embedded small clusters, i.e., MRPT2 or a comparable method with a satisfactory basis set size, at tenable cost? It appears that the answer is no.

**Summary and Conclusions**

Our study presents a detailed benchmarking of the capped-DFET method using embedded NEVPT2 to describe a complex electronic defect in a crystal, specifically, the negatively charged "NV center" in diamond. We found that using capped-DFET with emb-NEVPT2, we can reproduce VEEs and the ISC energy for spin triplet and singlet $N_CV_C^-$ with errors less than 0.1 eV. Furthermore, we show that although we derive an embedding potential from models subject to PBCs, which suffer from slowly vanishing charged defect - charge defect Coulomb interactions as a function of cell size, our embedded capped-cluster VEEs are free from such interactions, yielding predictions that are converged even for $V_{emb}$ derived for a 63-atom defective supercell. Therefore, we show that unlike other embedding methods, capped-DFET exhibits robust prediction that is approximately independent of the embedded cluster size, making it especially useful in reducing computational cost, as it enables use of medium supercell sizes to predict properties of defects at the dilute limit, with the caveat that the defect must be electronically localized within a few atoms. Bare large cluster sizes that remain computationally viable with our method of choice are unable to reproduce the embedded smaller cluster predictions, which highlights the advantage of embedding over purely cluster type models. With our tests, we confirm and thus establish an accuracy-retaining method that will potentially enable fully ab initio QM characterization of an array of localized defects in diamond that may be key materials as building blocks for quantum devices.

**Author Contributions**

JMPM conceptualized, generated and interpreted the data, and wrote the manuscript.

**Acknowledgements**

The research described in this paper was conducted under the Laboratory Directed Research and Development (LDRD) Program at Princeton Plasma Physics Laboratory, a national laboratory operated by Princeton University for the U.S. Department of Energy under Prime Contract No. DE-AC02-09CH11466. The United States Government retains a non-exclusive, paid-up, irrevocable, world-wide license to publish or reproduce the published form of this manuscript, or allow others to do so, for United States Government purposes. The simulations presented in this article were performed on Princeton University's Tiger HPC and Princeton Plasma Physics Laboratory's Stellar HPC resources managed and supported by Princeton Research Computing, a consortium of groups including the Princeton Institute for Computational Science and Engineering (PICSciE) and the Office of Information Technology's High Performance Computing Center and Visualization Laboratory at Princeton University. We thank Prof. Emily A. Carter, Dr. Phillips Hutchinson, and Dr. Vidushi Sharma for reviewing the first version of the manuscript.

**Competing interests**

The author declares no financial or non-financial competing interests

**Supporting Information**

Supplementary Methods, calculated optimized cell parameters for the three supercell sizes, defect-defect Coulomb interaction analysis, summary of basis set sizes, tabulated VEEs and excited state ISC energy for different cluster sizes, pure diamond DOS, charge density ($q=-1 - q=0$) and spin density difference plots, defective diamond DOS for different cell sizes, deference between DFT-HSE06- and DFT-$r^2$-SCAN-L-derived embedding potentials, state-averaged natural orbitals for (6e,8o) AS for $S = 0$, state-averaged natural orbitals for (10e,12o) and (12e,14o) ASs for $S = 0$, charge density difference (DFT-HSE06 – DFT-$r^2$-SCAN-L) plot, partitioning and embedding potential for $C_{126}N^-$ supercell for two different capped clusters.